\documentclass{emulateapj}
\usepackage{graphicx}
\usepackage{natbib}
\input{psfig.sty}
\def\msun{\rm {M}_{\odot}}

\def\simlt{\mathrel{\rlap{\lower 3pt\hbox{$\sim$}}\raise 2.0pt\hbox{$<$}}} 
\def\simgt{\mathrel{\rlap{\lower 3pt\hbox{$\sim$}} \raise 2.0pt\hbox{$>$}}} 
\def\lsim{\mathrel{\rlap{\lower 3pt\hbox{$\sim$}}\raise 2.0pt\hbox{$<$}}} 
\def\gsim{\mathrel{\rlap{\lower 3pt\hbox{$\sim$}} \raise 2.0pt\hbox{$>$}}} 
\def\lta{\mathrel{\rlap{\lower 3pt\hbox{$=$}}\raise 2.0pt\hbox{$<$}}} 
\def\gta{\mathrel{\rlap{\lower 3pt\hbox{$=$}} \raise 2.0pt\hbox{$>$}}} 
 
\newcommand{\beq}{\begin{equation}}
\newcommand{\eeq}{\end{equation}}
\shorttitle{Formation of massive black hole seeds}
\shortauthors{Devecchi \& Volonteri}

\begin{document} 
 
\title{Formation of the first nuclear clusters and massive black holes
  at high redshift} \author{B.~ Devecchi} \affil{Dipartimento di
  Fisica G.~Occhialini, Universit\`a degli Studi di Milano Bicocca,
  Piazza della Scienza 3, 20126 Milano,
  Italy}\email{bernadetta.devecchi@mib.infn.it} \and
\author{M.~Volonteri} \affil{Astronomy Department, University of
  Michigan, 500 Church Street, Ann Arbor, MI, 48109, USA }

 
\begin{abstract} 
 We present a model for the formation of massive black holes ($\sim
 1000 \msun$) due to stellar-dynamical processes in the first stellar
 clusters formed at early cosmic times ($z\sim10-20$).  These black
 holes are likely candidates as seeds for the supermassive black holes
 detected in quasars and nearby quiescent galaxies.  The high redshift
 black hole seeds form as a result of multiple successive
 instabilities that occur in low metallicity ($Z\sim 10^{-5}Z_\odot$)
 protogalaxies.  We focus on relatively massive halos at high redshift
 ($T_{\rm vir}\,>\,10^4$ K, $z\gsim 10$) after the very first stars in
 the Universe have completed their evolution.  This set of assumptions
 ensures that (i) atomic hydrogen cooling can contribute to the gas
 cooling process, (ii) a UV field has been created by the first stars,
 and (iii) the gas inside the halo has been mildly polluted by the
 first metals.  The second condition implies that at low density $H_2$
 is dissociated and does not contribute to cooling.  The third
 condition sets a minimum threshold density for fragmentation, so that
 stars form efficiently only in the very inner core of the
 protogalaxy.  Within this core, very compact stellar clusters
 form. The typical star cluster masses are of order $10^5\msun$ and
 the typical half mass radii $\sim 1$ pc. A large fraction of these
 very dense clusters undergo core collapse before stars are able to
 complete stellar evolution.  Runaway star-star collisions eventually
 lead to the formation of a very massive star, leaving behind a
 massive black hole remnant.  Clusters unstable to runaway collisions
 are always the first, less massive ones that form.  As the
 metallicity of the Universe increases, the critical density for
 fragmentation decreases and stars start to form in the entire
 protogalactic disk so that i) accretion of gas in the centre is no
 longer efficient and ii) the core collapse timescale increases.
 Typically a fraction $\sim 0.05$ of protogalaxies at $z\sim10-20$
 form black hole seeds, with masses $\sim 1000-2000 \msun$, leading to
 a mass density in seeds of a few $\simeq10^2\msun/{\rm
   Mpc}^{-3}$. This density allows enough room for black hole growth
 by accretion during the quasar epoch.  \end{abstract}
 

\section{Introduction} 
Supermassive black holes (BHs) are routinely detected in the centre of
galaxies, both in nearby quiescent galaxies, and as the engines that
power quasars and active galactic nuclei.  Observationally, we can
trace quasars until very high redshifts. Luminous quasars are detected
in the Sloan survey (e.g. Fan et al. 2001) at $z\simeq 6$,
corresponding to a time when the Universe was not even one billion
years old. This implies that the first BH seeds must have formed
earlier on. The currently favored scenario for BH seed formation
relies on the remnants of the very first generation of metal-free
stars (Population III, PopIII).  The first stars are believed to form
at $z\sim 20$ in halos which represent rare high peaks of the
primordial density field.  Simulations of the fragmentation of
zero-metallicity protogalaxies suggest a very top-heavy initial
stellar mass function, and in particular the production of very
massive stars with mass $>100 \msun$. PopIII stars in the mass range
$140\le m_\star\le 260\,\msun$ are predicted to make pair-instability
supernovae.  If zero metallicity very massive stars form above 260
$\msun$, they will rapidly collapse to BHs with little mass loss
(Fryer et al. 2001), leaving behind BHs with masses $ \simeq
10^2\,\msun$ (Madau \& Rees 2001).

Although this path to BH seed formation seems very natural, large
uncertainties exist on the final mass of PopIII stars. Even recent
simulations (Gao et al. 2006) have not clarified if PopIII stars are
indeed very massive, and in particular if they are above the threshold
($\simeq 260\msun$) for BH formation (but see Freese et al. 2008,
Spolyar et al. 2008, Natarajan et al. 2008). Furthermore, metal
abundances in extremely metal poor Galactic halo stars, which are
commonly thought to trace the enrichment products of the first
generation of stars, are incompatible with the yield patterns of zero
metallicity very massive stars ($\simgt100 \msun$, Tumlinson et
al. 2004).

Alternative routes to BH seed formation have been explored (Haehnelt
\& Rees 1993, Loeb \& Rasio 1994, Eisenstein \& Loeb 1995, Bromm \&
Loeb 2003, Koushiappas et al. 2004, Begelman, Volonteri \& Rees 2006,
Lodato \& Natarajan 2006), typically exploiting gas-dynamical
processes in metal-free galaxies. Gravitational instabilities can indeed lead
to a vigorous gas inflow into the very central region, supplying the
necessary matter for the formation of BH seeds.  The typical
conditions that lead to efficient gas infall and BH seed formation can
be summarized as (i) the host is massive enough that the virial
temperature exceeds $T_{\rm vir}\,>\,10^4$ K, so that gas is able to
cool and collapse via atomic hydrogen cooling. (ii) Molecular hydrogen
does not form as the gas cools and condenses. (iii) The gas has
primordial composition, so that metal line cooling is non-existent.

However, massive halos ($T_{\rm vir}\,>\,10^4$K, masses $\sim 10^7\msun$) are likely built-up from
mini-halos ($T_{\rm vir}\,<\,10^4$ K) that had collapsed earlier on.
Some of these halos might have experienced PopIII star formation, and
should be enriched with at least some trace amount of metals.
Fragmentation and formation of low mass stars starts as soon as gas is
polluted by metals created in the first PopIII stars. Efficient gas
collapse, leading to BH seed formation, is mutually exclusive with
star formation, as competition for the gas supply limits the mass
available.

However this first episode of efficient star formation can 
foster the formation of very compact nuclear star clusters 
(Clark et al. 2008, Schneider et al. 2006) where star
collisions can lead to the formation of a supermassive star, possibly
leaving a BH remnant with mass in the range
$\sim\,10^2-10^4\,M_{\odot}$ (Omukai et al. 2008).

The possibility that an ``intermediate-mass" BH could in principle
form as a result of dynamical interactions in dense stellar systems is
a long standing idea (Begelman and Rees 1978; Freitag et al. 2006b,a,
Ebisuzaki et al. 2001; Portegies Zwart and McMillan 2002; Miller \&
Hamilton 2002, Gurkan et al. 2004).  During their lifetime collisional
stellar systems evolve as a result of dynamical interactions. In an
equal mass system the central cluster core initially contracts as the
system attempts to reach a state of thermal equilibrium: energy
conservation leads to a decrease in the core radius as evaporation of
the less bound stars proceeds. As a result the central density
increases and the central relaxation time decreases. The core then
decouples thermally from the outer region of the cluster. Core
collapse then speeds up as it is driven by energy transfer from the
central denser region (Spitzer 1987).
This phenomenon is greatly enhanced in multi-mass systems like
realistic star clusters. In this case the gravothermal collapse
happens on a shorter timescale as dynamical friction causes the more
massive stars of mass $m$, to segregate in the centre on a time-scale
$t_{\rm df}=(\langle m\rangle/m)\,t_{rh}$ (where $t_{rh}$ is the half
mass relaxation timescale and $\langle m\rangle$ is the mean stellar
mass in the cluster). If mass segregation sets in before the more
massive stars evolve out of the main sequence ($\sim$ 3 Myr), then a
sub-system decoupled from the rest of the cluster can form, where
star-star collisions can take place in a runaway fashion that
ultimately lead to the growth of a very massive star (VMS) (Portegies
Zwart et al. 1999).

Yungelson et al. (2008) study the fate of \textit{solar composition}
VMSs in the mass range 60-1000 $M_\odot$. They find that all VMSs are
likely to shed most of their mass via winds well before experiencing a
supernova explosion. Solar composition VMSs are therefore expected to
end their lives as objects less massive than $\sim 150 M_\odot$,
collapsing into BHs with mass $\lsim 150 M_\odot$ or exploding as
pair-instability supernovae. The growth of a VMS should be much more
efficient at \textit{low metallicity}.  Low metallicity can modify the
picture in different ways.  First, 
at sub-solar (but still not primordial) metallicity, all stars with
masses $\gta40M_\odot$ are thought to collapse directly into a black
hole (Heger et al. 2003) without exploding as supernovae.  Second, the
mass loss due to winds is much more reduced in metal-poor stars, which
greatly helps in increasing the mass of the final remnant.

In this paper we investigate the formation of BHs, remnants of VMS
formed via stellar collisions in the very first stellar clusters at
early cosmic times.  We derive the properties of the ensuing BH
population, that can represent the ``seeds" of the supermassive BHs
that we observe in today's galaxies.  In the following we summarize
the main features of our model:
\begin{itemize}
\item we focus on halos with virial temperatures $T_{vir}\simgt10^4$ K at $z>5$ {\it after}  the first episode of star formation, hence with a low , but non-zero, metallicity
\item low metallicity gas can fragment and form low-mass stars only if the gas density is above a certain threshold, $n_{crit,Z}$
\item halos possess angular momentum, acquired through interaction with their neighbors, hence gas collapse ultimately leads to the formation of a disk
\item if the forming disk is Toomre-unstable, instabilities lead to mass infall instead of fragmentation into bound
clumps and global star formation in the entire disk
\item the gas inflow increases the central density, and within a certain, compact, region  $n>n_{crit,Z}$
\item star formation ensues and a dense star cluster is formed
\item if the star cluster goes into core collapse in $\simlt$ 3 Myr runaway collisions of star form a supermassive star, leading to a massive BH remnant.
\end{itemize}
We describe the physical mechanism for cluster formation in section 2,
where we discuss the conditions leading to very compact stellar
clusters in low metallicity ($Z\sim 10^{-5}Z_\odot$) protogalaxies.
We discuss dynamical instabilities (Section 2.1) driving gas towards
the centre, and fragmentation occurring at high densities (Sections
2.2 and 2.3).  The properties of the star clusters, VMSs and BHs are
described in Sections 2.4 and 2.5. We present our results in Section 3
and discuss the implications of our model in Section 4.

\section{BH seed formation:  model set-up}
We start by considering a halo of mass $M_h$, with virial circular
velocity $V_h$, virial radius $R_h$, angular momentum $J$. The
rotational support can be quantified in terms of the spin parameter
$\lambda \equiv J |E|^{1/2}/G M_h^{5/2}$, where $E$ is the total
energy of the halo. We further assume that a fraction $m_d\sim 0.05$
of the halo mass is in a gaseous baryonic component than can cool and
condense.  The collapsing mass will then be $m_d\,M_h$.  The angular
momentum of the collapsing baryons will be a fraction $j_d$ of the halo
angular momentum, with $j_d/m_d=1$ if the specific angular momentum of
baryons is conserved during collapse.  In this work we focus on
relatively massive halos ($T_{\rm vir}\,>\,10^4$ K, ensuring that gas
can cool by atomic hydrogen) after the birth of the first stars in the
Universe.   $10^4$ K halos are likely build up by smaller halos that
  have already experienced a first episode of star formation. We
  assume that the first PopIII stars have been able to affect the gas
  both radiatively (thus precluding subsequent $H_2$ cooling in the
  halo) and chemically.  The first condition implies that, at least
at the low density at which the collapse starts, $H_2$ is dissociated
and does not contribute to the cooling of the gas (see also Figures 9
and 10 in O'Shea \& Norman 2008).  The second condition ensures that
metals (and dust) in small quantities can cool down the gas
efficiently only as the gas density reaches a critical threshold (see
Section 2.2, and Figures 2, 4 and 5 in Smith et al. 2008). Jointly,
the two conditions ensure that at first cooling is driven by atomic
hydrogen only.

\subsection{Disk formation and  mass inflow}
We describe the early evolution of the cooling and collapsing baryons
in a simple way, assuming adiabatic response of the halo to gas
cooling and disk formation (e.g., Mo, Mao \& White 1998, Oh \& Haiman
2002, Lodato \& Natarajan 2006). As we are assuming negligible $H_2$,
the tenuous gas cools down by atomic hydrogen only until it reaches
$T_{\rm gas}\sim 4000$ K.  At this point the cooling function of the
atomic hydrogen drops by a few orders of magnitude, and contraction
proceeds nearly adiabatically.

Given the presence of angular momentum, the contraction in the
equatorial plane stops as the gas becomes rotationally supported, and
the collapse ultimately leads to the formation of a disk. As pointed
out by Mo, Mao $\&$ White (1998), the properties of the disk can be
determined by assuming a profile for the surface density $\Sigma$. As
the mass assembles in the protodisk the surface density increases. If
for the final configuration the Toomre parameter: \beq Q=\frac{c_s
  \kappa}{\pi\,G\,\Sigma}<Q_c, \eeq where $c_s$ is the sound speed,
$\kappa$ is the epicyclic frequency and $Q_c$ is a critical value,
then the disk becomes unstable, and it develops bar-like
structures. We argue that if the destabilisation of the system is not
too violent, instabilities lead to mass infall instead of
fragmentation into bound clumps and global star formation in the
entire disk (Shloshman et al. 1990, Lodato \& Natarajan 2006).  This
is the case if the inflow rate is below a critical threshold
$\dot{M}_{ max}=2\alpha_{c}\frac{c^3_s}{G}$ the disk is able to
sustain (where $\alpha_c\,\sim\, 0.12$ describes the viscosity) and
molecular and metal cooling are not important. We substantiate further
our argument in Section 2.2.  In the following, we discuss what
happens to the disk if fragmentation is (at least initially)
suppressed.

The classical analysis yields $Q_c\simeq1$ for axi-symmetric
disturbances. However a disk can develop large scale non-axisymmetric
instabilities even for $Q>1$ so that a limit value $Q_c\sim 2$ is
probably a better guess (Lodato 2008).  We will assume in the
following $Q_c=2$. The frequency of disk instabilities as a function
of $Q_c$ is discussed in Lodato \& Natarajan (2007) and in Volonteri,
Lodato \& Natarajan (2008), who showed that any $Q_c<1.5$
underestimates the BH population, even assuming a 100\% of BH
formation in unstable disks.

The torques induced by self-gravity can cause a relatively fast
redistribution of gas within the disk: the gas shocks and loses
angular momentum thus sinking to the centre of the system. As gas
flows into the central region, the surface density in the outer part
of the disk decreases until the Toomre parameter nears the critical
value once again. At this point the infall stops. In this sense the
instabilities described here are self-regulating and end up in a
condition of marginal stability. Ultimately, the mass routed in the
central part corresponds to the amount necessary for the outer disk to
be marginally stable. Let us define this central mass condensation as
a fraction $m_a$ of the halo mass, that  corresponds to a baryonic
  mass $m_a M_h$.

A marginally stable, isothermal disk has the surface density profile
consistent with a Mestel disk, with
$\Sigma=\Sigma_0\left(\frac{R}{R_0}\right)^{-1}$. This is our initial
condition for the outer region of the disk.  As the mass $m_aM_h$ is
added to the mass already in place in the centre, the inner surface
density profile steepens, $\Sigma\propto \left(\frac{R}{R_0}\right)^{-\gamma}$, with $\gamma>1$ . 
Mineshige $\&$ Umemura (1997) studied the collapse of an
isothermal, rotationally supported, self-gravitating collapsing disk. 
The collapsing disk develops a different structure in the inner and outer part, 
and the collapse proceeds with time as follows:

\begin{equation}
  \Sigma(R)\sim \left\{
\begin{array}{lr}
\Sigma_{in}\left(\frac{R}{R_0}\right)^{-5/3}\left(\frac{t}{t_0}\right)^{2/3}
&R<R_{tr}\\ 
\Sigma_0\left(\frac{R}{R_0}\right)^{-1} & R>R_{tr}\\
\end{array}
\right.
\label{eq:mine}
\end{equation}
\noindent
where the transition radius $R_{tr}$ between the two profiles evolves
linearly with time. As the transport associated with the development
of gravitational instabilities can be well described in terms of an
effective viscosity (Lodato \& Natarajan 2006), we take the result of 
Mineshige $\&$ Umemura as a reference for our system and we assume 
that the inner profile steepens with $\gamma=5/3$ unless otherwise noted.

The final configuration of the structure would then be characterized
by an outer surface density profile of a Mestel disk and a central
denser region:
\begin{equation}
  \Sigma(R)\sim \left\{
\begin{array}{lr}
\Sigma_{in}\left(\frac{R}{R_0}\right)^{-\gamma}
&R<R_{tr}\\ 
\Sigma_0\left(\frac{R}{R_0}\right)^{-1} & R>R_{tr}\\
\end{array}
\right.
\label{eq:mine_new}
\end{equation}
\noindent
where $\Sigma_{in}$ can be written as $\Sigma_0(R_0/R_{tr})^{1-\gamma}$ by imposing
continuity at $R_{tr}$.  At this point the parameters of the final
disk are $\Sigma_0$, $R_0$, $R_{tr}$ and the fractional mass of the
halo $m_a$ that participate to the infall. They can be determined by
imposing the conditions of mass and angular momentum conservation,
$Q=Q_c$ (for $R>R_{tr}$) and by imposing that a mass $m_aM_h$ is added
to that already in place inside $R_{tr}$.  For simplicity we assume
that the dark matter halo hosting the protogalaxy follows an
isothermal profile. The resulting disk properties are as follows:

\beq
\Sigma_0=\frac{10m_d(m_d/j_d)^2\,H_z\,V_h(1-m_a/m_d)^3}{16 \pi G \lambda^2},
\eeq
\beq
 R_0=\frac{2\sqrt{2}(j_d/m_d)\lambda R_h}{(1-m_a/m_d)}           
\eeq

\begin{center}
  \begin{equation}
    m_a=m_d\left[1-\sqrt{\frac{8\lambda}{m_dQ_c}\frac{j_d}{m_d}\left(\frac{T_{gas}}{T_{vir}}\right)^{1/2}}\right]
    \label{eq:ma}
  \end{equation}
\end{center}

\begin{center}
  \begin{equation}
    R_{tr}=\frac{m_aM_h}{4\pi\Sigma_0 R_0},
    \label{eq:rtr}
  \end{equation}
\end{center}

where $H_z$ is the Hubble constant at redshift $z$. 
The vertical structure of the disk is determined by solving the
equation for hydrostatic equilibrium.  For a disk that is isothermal
and self-gravitating, this implies a z-dependency for the density
$\propto \cosh^{-2}(h/H)$ where $H$ is the vertical scale-height and
it is equal to $H(R)=c^2_s/\pi G\Sigma(R)$. 
We can therefore express the inner disk density as:
\beq
n(R,h)=n_0  \left( \frac{R} {R_0} \right)^{-5/3}\cosh ^{-2}(h/H(R)),
\label{eq:density}
\eeq where $n_0$ is a function of the disk parameters, $\Sigma_0$,
$R_0$ and $\gamma$ (see Appendix A).

\subsection{Fragmentation of the disk}

A necessary condition for the inflow process described in the previous
Section is that star formation does not take place in the
\textit{entire} disk.  If this happens, the gas that would otherwise
flow into the central region, is consumed as it is converted into
stars.

Rice et al. (2005) proposed that fragmentation in a thin disk sets in
when the gravitationally induced stress exceeds a critical value.
Describing angular momentum transport in terms of the $\alpha$
prescription for viscous dissipation (i.e. that the torque strength
can be expressed in terms of $\alpha$, Shakura \& Sunyaev 1973), then
the critical threshold for fragmentation $\alpha_c$ determines how
much angular momentum can be transported in a steady state. In this
sense the fragmentation boundary is ultimately due to the inability of
the disk to redistribute angular momentum on a sufficiently short
timescale. In the classical analysis (Gammie 1996) fragmentation
develops as the condition $t_{cool}=t_{dyn}$ is reached. In the
framework of Rice et al. (2005) we can expect that the same effect
develops if the mass inflow from the halo induces too strong a
stress. Lodato \& Natarajan (2006) apply this framework to disks of primordial composition. By
requiring the mass-accretion rate from the halo
$\dot{M}_h=m_d\frac{V^3_h}{G}$ to be less than the maximum value
$\dot{M}_{ max}=2\alpha_{c}\frac{c^3_s}{G}$ the disk is able to
sustain, they argue that to avoid fragmentation it must be

\begin{center}
  \begin{equation}
    \frac{T_{vir}}{T_{gas}}<\left[\frac{4\alpha_c}{m_d}\frac{1}{1+m_a/m_d}\right]^{2/3}.
  \label{eq:frag}
  \end{equation}
\end{center}
\noindent
where  $\alpha_c\,\sim\, 0.12$ (Clarke et al. 2007) is the
critical value for fragmentation.

We assume $T_{gas}$ = 4000 K, which corresponds to $\dot{M}_{max}\sim
10^{-2}$M$_{\odot}$ yr$^{-1}$. The joint conditions, $m_a>0$ (Equation~\ref{eq:ma}) and
Equation~\ref{eq:frag} then impose an upper limit to the virial
temperature $T_{vir}\lta 1.8\times10^4$ K.  We therefore consider
halos with $10^4$ K$ \lta T_{vir}\lta1.8\times10^4$ K.

By considering these gas and virial temperatures we can ensure that the fragmentation threshold is not
attained at least at $R>R_{tr}$.  However, $T_{gas}=4000$ K requires
the absence of coolants other than atomic hydrogen. If $H_2$
contributes efficiently as a coolant, the gas temperature drops, and
Equation~\ref{eq:frag} implies that massive halos are subject to 
strong fragmentation. For instance, if gas cools to
$T_{gas}\sim 200$ K, only halos with $T_{vir}\,\lta\,1000$ K satisfy
Equation~\ref{eq:frag}. At redshift $\sim$ 20 this corresponds to halo
masses around a few $10^5 \msun$.

The issue of $H_2$ cooling suppression in the presence of a strong UV
field has been addressed in a number of recent studies on the first
galaxies.  Even if a strong UV background is not already in
place, a ``local'' UV field can be established by the presence of
PopIIIs in surrounding halos. Eventually, $H_2$ can be suppressed by
previous PopIII star formation in the smaller systems assembled to
form the final $T_{vir}\sim 10^4$ K halo (see Ciardi (2008) or Ciardi
\& Ferrara (2005) for a discussion on this issue). O'Shea \& Norman
(2008) have shown that halos with $T_{vir}\sim 10^4$ K embedded in a
UV background with strength higher than $\sim 3\times10^{-23}$erg
s$^{-1}$ cm$^{-2}$ Hz$^{-1}$ sr$^{-1}$ are not subject to efficient
cooling and fragmentation in the disk (see their Fig. 9b) and it is
only in the central region (${\rm r}< 10$ pc) that $T_{gas}$ drops
down to $\sim 10^3$ K, leading to the formation of a central single
massive star born in isolation.  Halos subject to a weaker UV field
show the same radial behavior but shifted at lower temperature. If
$H_2$ cooling in fact does not act on large scales, it does not affect
the inflow process described in Section 2.1 so that material can still
be transported to small distances by dynamical instabilities.

In metal-free conditions, the gaseous density in the central region
increases until H$_2$ is activated (O'Shea \& Norman 2008, Omukai et
al. 2008, and references therein) and the formation of a Pop III star
can proceed.  On the other hand, if the gas has been enriched to a
certain level, fragmentation can take place and an entire stellar
cluster is formed (Clark et al. 2008). It is this last situation that
we want to examine in more details, i.e. the possibility that a
cluster of stars is formed instead of a single massive star.

\subsection{Critical metallicity for fragmentation}

Various authors have suggested that the presence of a certain amount
of metals is the key ingredient in order to produce efficient
fragmentation (Schneider et al. 2006; Bromm et al. 1999, 2002; Omukai et al. 2008; Clark et
al. 2008 and references therein). Let us define a critical metallicity
$Z_{crit}$ where transition from PopIII to ``normal" star formation
occurs.  $Z_{crit}$ depends on the coolants at work.
Bromm et al. (2001) simulated the collapse of a halo of $2\times 10^6
M_{\odot}$ at different values of $Z$.  They estimate 
$Z_{crit}\sim10^{-3.5}$ and show that  when $Z>Z_{crit}$ a 
rotationally supported disk that fragments vigorously can form. More recently, Clark
et al. (2008) simulate the collapse of a rotating cloud polluted by
dust. They show that a tightly packed cluster of protostars is formed
in the centre. 

To estimate the value of $Z_{crit}$ at which gas start to fragment one
can compare the cooling rate $\Lambda_{cool}$ to the adiabatic heating
rate $\Gamma_{ad}$. Fragmentation requires the $\Lambda_{cool} \gta
\Gamma_{ad}$. Once this happens, the temperature starts to decrease as
the density increases until cooling stops to be efficient. The
condition $\Gamma_{ad}<\Lambda_{cool}$ in fact constraints the
metallicity at a given temperature and density to be greater than a
critical value $Z_{crit}$. For an initially isothermal gas this
condition provides a relationship between metallicity and
density. Equivalently, if the gas is characterized by a fixed
metallicity Z, then only those regions with density greater than a
given threshold $n_{crit,Z}$ are able to cool down efficiently and
fragment into stars. This is the condition that we apply to collapsing
disks to determine if they can develop into stellar clusters.

Santoro et al. (2006) studied the condition for fragmentation in a low
metallicity Universe assuming that PopIII stars are the main sources
of pollution. The $Z_{crit}-n_{crit,Z}$ relation depends on the ratio
between the different species of coolants (see Figure 9 in Santoro et
al. 2006).  In all cases, the critical metallicity increases with
decreasing density untill $n$ reaches the critical value for
collisional de-excitation of the dominant coolants.  We adopt as a
reference the curve corresponding to PopIII stars in the intermediate
mass range (mass range 185--205$\msun$; however the solar abundance
ratio produces a similar pattern as well, cfr. solid and dotted curves
in Figure~10 in Santoro \& Shull 2006).  We discuss how our results
depend on the specific $Z_{crit}-n_{crit,Z}$ relation in section
3.2.2.

The metallicity of gas in a given halo most probably depends on its
mass and on the redshift.  The cosmic metallicity history (MEH) has
been investigated both from an observational and a theoretical
prospective (Scannapieco et al. 2003, Tornatore et al. 2006, Savaglio
2006, Savaglio et al. 2005, Prochaska et al. 2003, Prochaska et
al. 2007, Kulkarni et al. 2005, Li 2007). The observed mean
metallicity decreases with redshift, but the rate of decrease depends
on the type of sources studied. From studies of QSOs-DLAs,
$Z/Z_{\odot}$ scales as $10^{-\beta\,z}$ with $\beta=0.36$ (Li 2007),
while measurement based on GRBs point towards a shallower decline, so
that at $z\sim3-4$ already $Z=10^{-1} Z_{\odot}$ (Savaglio 2006). The
differences in the slope and in the normalization are often ascribed
to different histories of metal enrichment in QSOs and GRBs hosts (Li
2007 and references therein).

We model the MEH based on fits presented in Li 2007, that we refer to
for additional details.  In our reference model we adopt $\beta=0.36$,
that is $Z/Z_{\odot}=0.35 \times 10^{-0.36z}$.  The case $\beta=0.18$,
leading to $Z/Z_{\odot}=0.63 \times 10^{-0.18z}$, is discussed in
section 3.2.3.  We further allow a logarithmically uniform scatter in
$Z$ of $\Delta \log(Z)=1.5$, based on the observed scatter in $Z/Z_{\odot}$
from measurements of the QSO-DLAs.  We wish to stress that our
treatment of MEH is highly simplified. In a forthcoming paper we will
determine self-consistently the evolution of the MEH from stellar
winds pollution.

\subsection{Stellar cluster formation}
The central star cluster forms within the region where gas at a given
metallicity has reached the critical density for fragmentation.  Given
a specific density profile (Equation~\ref{eq:mine}) the requirement
$n>n_{crit,Z}$ translates into a condition $R<R_{SF}$, where we define
$R_{SF}$ as the radius where the density in the plane $h=0$, reaches
the value $n_{crit,Z}$ for the gas metallicity $Z$, that is: 
\beq
n(R=R_{SF},h=0)=n_{crit,Z}.
\label{eq:R_SF}
\eeq 

As long as $n(R_{tr},0)<n_{crit,Z}$, $n(R,h)$ is defined by
Equation~\ref{eq:density} and the radius within which gas fragments
and stars form can be written as:
\begin{equation}
R_{SF}=R_{tr}\left[\frac{\Sigma_0}{c_s}\frac{R_0}{R_{tr}}\sqrt{\frac{\pi
      G}{2\mu m_H n_{crit,Z}}}\right]^{1/\gamma}
\label{eq:RSF}
\end{equation}
\noindent
where the metallicity dependence is included implicitly in
$n_{crit,Z}$ (see Appendix A), $\mu$ is the molecular weight and $m_H$
is the proton mass.  The half mass radius can be expressed as a
function of the cluster radius in a very simple way, as 
\beq
R_h=2^{1/(\gamma-1)}\,R_{SF}.  
\eeq 
For a given $\gamma$ the mass in
stars of the cluster can be calculated as
\begin{eqnarray}
M_{cl}&=&\epsilon_{SF}2\pi\int^{R_{SF}}_0\Sigma(R)RdR\nonumber \\
&=&\epsilon_{SF}\left[\frac{2\pi}{2-\gamma}\frac{\Sigma^2_0R^2_0}{c_s}\sqrt{\frac{\pi
      G}{2\mu m_u
      n_{crit}}}\right]^{\frac{2-\gamma}{\gamma}}\left(\frac{m_aM_h}{\gamma-1}\right)^{\frac{2\gamma-2}{\gamma}}\, 
\end{eqnarray}
where $\epsilon_{SF}$ is the fraction of gas converted into stars. We
assume $\epsilon_{SF}=0.25$, consistently with the star formation
efficiency in the low redshift Universe (Lada \& Lada 2003).  Given
the metallicity, $Z$, it is now possible to determine the extent of
the region inside which star formation is allowed.  The cluster
properties for a given halo, are therefore uniquely described for any
given metallicity.

At high metallicities, the case $n(R_{tr},0)>n_{crit,Z}$ becomes
common.  This implies $R_{SF}>R_{tr}$, that is, star formation takes
place in the region of the disk where no inflow is taking place. In
other words, once the gas reaches $R_{SF}$, it is no longer able to be
routed efficiently in the inner region as star formation starts to
consume gas. In the remainder of the paper we will assume
conservatively that if $R_{SF}>R_{tr}$ no cluster formation occurs, as
stars are formed in the disk rather than in the central compact
region.

\subsection{Runaway instability of the central cluster}
The conditions under which ``mass segregation instability'' can occur
have been investigated in a series of papers, focusing on clusters in
the present-day Universe.  A successful core collapse requires that
the core collapse time be less than the main sequence lifetime of the
most massive stars (mass losses from supernovae expand the core and
increase interaction times). The main sequence lifetime of massive
stars asymptotes to about 2.5 Myr because all stars go off the main
sequence when they have consumed about 15\% of their hydrogen, and for
high-mass stars with luminosities approaching Eddington, $L\propto M$
(not $M^{3.5}$ as is the case for lower masses) and then the lifetime
$\propto 0.15M/L\sim$ const.

With a Monte Carlo code, G{\"u}rkan et
al. (2004) found that typically the mass of the collapsing core is
$10^{-3}$ times that of the entire cluster. Similarly, Portegies Zwart
and McMillan (2002) related the mass of the VMS with the parameters of
the cluster taking into account both numerical simulations and
analytical arguments. Portegies Zwart \& McMillan (2002) have shown
that core collapse occurs on a timescale

\begin{eqnarray}
t_{cc}&\simeq&
3{\rm Myr}\,\left(R_{\rm h}\over 1{\rm pc}\right)
^{3/2}\left(M_{cl} \over 5\times10^5\,M_\odot \right)^{1/2}\times \nonumber \\
& & \left(  10\,M_\odot  \over  \langle m \rangle \right) \left(\frac {8.5}{\ln \lambda_C}\right),
\label{eq:tcc}
\end{eqnarray}

where $\ln{\lambda_C}$ is the Coulomb logarithm of dynamical friction
(Binney \& Tremaine 1987).  The same simulations also find that the
mass of the VMS can reach values as high as $10^3\, M_{\odot}$ (see
also Freitag et al. 2006b).  The final mass of the VMS however depends
on complex phenomena related to both the dynamics and hydrodynamics
of the collisions. 
At solar metallicity the growth of a star with mass greater than $\sim
100 M_{\odot}$ can be highly problematic as mass loss occurs both
during the main sequence phase and at the end of the evolution when
the star collapses into a BH. The growth of a VMS should in principle
be much more efficient for metal-poor stars as in this case as mass
loss should be strongly reduced. Recent models of stellar evolution at
$Z\sim 10^{-5}\,Z_{\odot}$ have shown that mass loss due to stellar
winds during the main sequence phase is almost unimportant and that
the main contribution to the reduction of the stellar mass is due to
the effect of rotation. This can reduce the mass of the star by a
factor of order 2-4 (Meynet et al. 2008). If the
final mass achieved by the VMS is greater than $\sim 260\msun$, then
after the main sequence it collapses into a BH retaining most of its
mass (Heger et al. 2003).

To estimate of the final mass of the VMS we follow the treatment
outlined in Portegies Zwart (2002).  We assume that the mass of the BH
seed, $M_{BH}$, corresponds to the final mass of the
VMS: 
\begin{equation} M_{BH}=m_*+4\times10^{-3}M_{cl}f_c
  \ln{\lambda_C}\ln{\left(\frac{t_{MS}}{t_{cc}}\right)} \label{eq:BH}
\end{equation}

Here $t_{MS}\,=\,3$ Myr is the main sequence lifetimes of massive
($>\,40\,M_{\odot}$, Hirschi 2007) stars, $m_*$ is the initial mass of the seed star
that experiences runaway growth and $f_c$ is a factor used to
calibrate the analytical expectation with direct numerical
simulations.  Portegies Zwart \& McMillan (2002) find $f_c\,=\,0.2$
for $\ln(\lambda_C)\,=\ln(0.1M_{cl}/\langle m \rangle )$. We adopt the
same values as Portegies Zwart \& McMillan (2002) for both parameters,
further assuming $m_*=\langle m \rangle=10\msun$ consistent with the
characteristic stellar mass in $10^{7-8} \msun$ halos at redshift
$\sim 10$ (see Figures 2 and 3 in Clarke \& Bromm 2003).

\begin{figure} 
\includegraphics[width=\columnwidth]{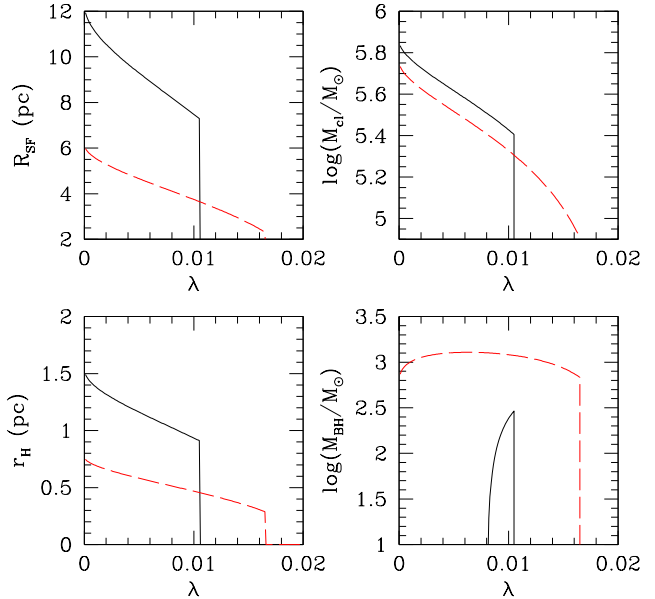} 
\caption{Cluster properties as a function of the spin parameter
  $\lambda\equiv J |E|^{1/2}/G M_h^{5/2}$ (i.e., the fraction of halo
  support given by rotation) for 2 critical densities for
  fragmentation, $n_{crit}=10^{3}$ cm$^{-3}$ (solid line), and
  $n_{crit}=10^{4}$ cm$^{-3}$ (dashed line). Curves are truncated when
  $R_{SF}$ equals $R_{tr}$.}
\label{fig:fig1} 
\end{figure}

It is important to stress that Equation~\ref{eq:BH} provides an upper
limit to $M_{BH}$.  However, once the seed is born, the BH is still
embedded into a dense cluster of stars.  Even if its growth has been
limited before the collapse of the VMS, the remnant can still gain
mass by accretion of stars (for the growth rate of a massive BH hosted
in a cluster of stars see for example the models of Marchant \&
Shapiro 1979). If the combination of star formation and stellar
feedback of the cluster stars do not deplete of gas the inner region,
this surviving gas can supply an additional reservoir of material for
BH growth. Both this processes can contribute to bring the BH mass to
values as high as Equation~\ref{eq:BH} would suggest.

\begin{figure} 
\includegraphics[width= \columnwidth]{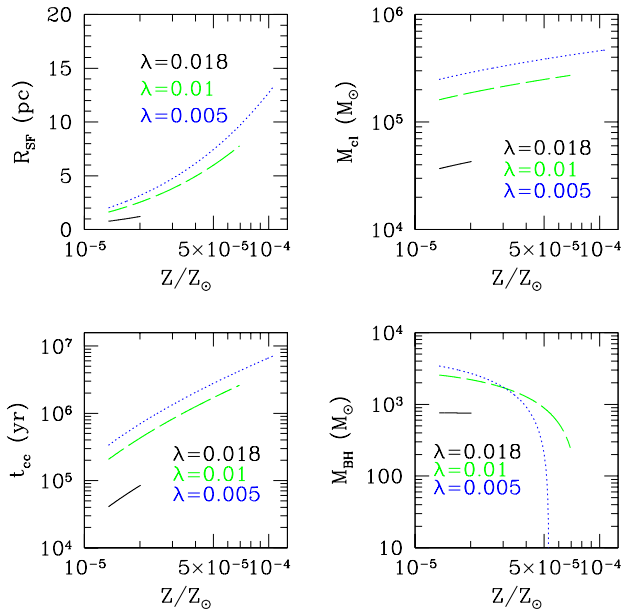} 
\caption{Cluster properties as a function of metallicity for 3 spin
  parameters: $\lambda=0.018$ (solid line), $\lambda=0.01$ (dashed
  line), $\lambda=0.015$ (dotted line). Each curve begins at low
  metallicity when halos starts star formation and ends when $R_{SF}$
  equals $R_{tr}$; this happens at different critical metallicities
  for different spin parameters. Here the $Z_{crit}-n_{crit,Z}$
  relationship follows the intermediate PopIII mass case in Santoro et
  al. 2006. }
\label{fig:fig2} 
\end{figure}

Clusters forming in unstable disks might have some degree of
rotation. In this case the gravogyro instability (Inagaki \& Hachisu
1978) could contribute to accelerate the core collapse. The gravogyro
instability is believed to occur in systems that exhibit a radial
gradient of the angular speed. In analogy with viscous transport,
angular momentum is transfered outwards. The core of the star cluster
then contracts because of a deficit in the centrifugal force. Ernst et
al. (2007) show that the gravogyro instability occurs in clusters with
equal-mass stars, but in systems with two-mass components the effect
of rotation seems negligible, as mass segregation and rotation compete
in leading the evolution of the stellar cluster. For this reason we
neglect its contribution in this work.

\subsection{Cluster Properties}
Various cluster properties: size (Equation 11), half-mass radius
(Equation 12), mass (Equation 13) are shown in Figure~\ref{fig:fig1}
as a function of the spin parameter for two representative values of
the critical density for fragmentation.  We also present BH masses
(Equation 15), where we have further imposed the condition $t_{cc}<3$
Myr.

As $\lambda$ increases, less mass inflows within $R_{tr}$, and
$R_{SF}$ is reached at smaller radii at a given $n_{crit}$. Therefore
$R_{SF}$ and $M_{cl}$ decrease with increasing $\lambda$, contrary to
the disk size $R_0$. Note how at low critical densities (corresponding
to high metallicities) only a very small fraction of systems can
undergo core collapse, as on the one hand cluster formation is
suppressed for large $\lambda$ (as $R_{SF}>R_{tr}$), on the other hand
clusters are too massive and large at small $\lambda$ for fulfilling
the condition $t_{cc}<3$Myr. Clusters undergoing core collapse have
preferentially large spin parameters, within the region where cluster
form. Figure~\ref{fig:fig2} shows cluster sizes, masses, core collapse
timescales (Equation 13) and BH masses as a function of metallicity.
As the metallicity increases, the critical density for fragmentation
decreases and cluster formation is eventually suppressed when
$n(R_{tr},0)>n_{crit,Z}$ ($R_{SF}>R_{tr}$).  At metallicities below
$10^{-5}\,Z_\odot$ fragmentation is impossible for the specific choice
of the $Z_{crit}-n_{crit,Z}$ relationship (intermediate PopIII mass
case in Santoro et al. 2006).

\begin{figure} 
\includegraphics[width=\columnwidth]{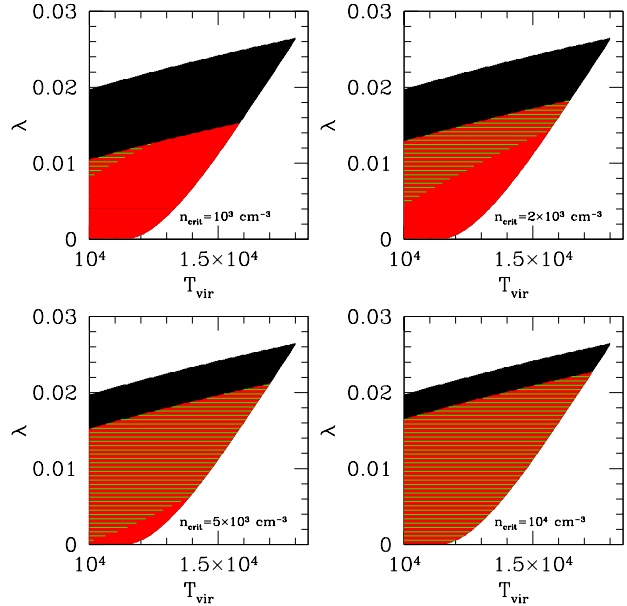}
\caption{Parameter space (virial temperature, spin parameter) for
  cluster and BH formation. Here we select halos with $T_{vir}>10^4$ K
  at $z=12$, and derive the disk and cluster properties, assuming a
  single critical density for fragmentation (top-left:
  $10^4$cm$^{-3}$, top-right: $2\times10^4$cm$^{-3}$, bottom-left:
  $2.5\times10^4$cm$^{-3}$, bottom-right: $3\times10^4$cm$^{-3}$; for
  the chosen $Z_{crit}-n_{crit,Z}$ relationship these densities
  correspond respectively to
  $\log(Z/Z_\odot)$=--4.2;--4.3;--4.4;--4.5). The black shaded area
  shows the range of temperatures and spin parameters where disks are
  Toomre unstable and the joint conditions, $m_a>0$
  (Equation~\ref{eq:ma}) and Equation~\ref{eq:frag} are fulfilled. The
  lighter shaded area selects the systems where $R_{SF}<R_{tr}$.  The
  hatched area picks the subsample of clusters where $t_{cc}<3$Myr,
  where VMSs and BH seeds can form. }
\label{fig:fig3} 
\vspace{0.5cm} 
\end{figure}

The parameter space (virial temperature, spin parameter) where the
multiple instabilities are efficient is shown in
Figure~\ref{fig:fig3}. Here we select halos with $T_{vir}>10^4$ K at
$z=12$, and derive the disk and cluster properties, assuming a single
critical density for fragmentation (from $10^3$cm$^{-3}$ to
$10^4$cm$^{-3}$). The higher the critical density for fragmentation
(i.e., the lower the metallicity, see Section 2.3) the more compact
are the clusters, and the shorter is $t_{cc}$.  When $n_{crit,Z}<10^3$
cm$^{-3}$ no clusters can undergo core collapse in less than 3
Myr. When $n_{crit,Z}>10^4$ cm$^{-3}$ all forming clusters undergo
core collapse in less than 3 Myr.

\section{Results}
\subsection{Summary of the general procedure}
We first summarize the procedure taken in order to determine the
properties of the BH seed population. We calculate the mass of halos
that at redshift $z$ correspond to virial temperatures $10^4$ K$\lta
T_{vir}\lta 1.8\times10^4$ K (Barkana \& Loeb 2001) and we determine
their frequency using a modified version of the Press \& Schechter
formalism (Sheth \& Tormen 1999) in a WMAP5 cosmology (Dunkley et
al. 2008).  To each halo we assign a value of the spin parameter,
$\lambda$, extracted from the probability distribution found in the
Millennium simulations (Bett et al. 2007): 
\beq P(\log
\lambda)=A\left( \frac{\lambda}{\lambda_0} \right)^3 \exp \left[
  -\zeta \left( \frac{\lambda}{\lambda_0} \right)^{3/\zeta} \right],
\eeq 
where $\lambda_0=0.0043$ is the peak location, $\zeta=2.509$ and
the normalization reads $A=3\ln10\zeta^{\zeta-1}/\Gamma(\zeta)$,
with $\Gamma$ being the gamma function.  This set of assumptions
allows us to calculate the initial disk properties, $\Sigma_0$, $R_0$,
$Q_c$, $R_{tr}$ and $m_a$.

We assign to each halo a metallicity, $Z$ by extrapolating at higher
redshift the fit to the observational constraints of the MEH
($Z\propto 10^{-\beta z}$), taking also into account the observed
metallicity scatter. We then calculate $n_{crit,Z}$ from a given
$Z_{crit}-n_{crit,Z}$ relation.  If a protogalaxy has $Q<Q_c$ we
determine the properties of the stellar cluster ($R_{SF}$, $M_{cl}$
and $R_h$). We then check if the cluster can develop runaway
instability via Equation~\ref{eq:tcc}, and we select the systems where
$t_{cc}<3$Myr. For these unstable clusters we calculate the expected
mass of the seed BH from Equation~\ref{eq:BH}.  In Table \ref{tab:sim}
we summarize all the different cases we describe in the next sections.

\begin{table}[h]
  \begin{center}
    \begin{tabular}{|c|c|c|c|}
      \hline
      Run &   MEH  & $Z_{crit}-n_{crit,Z}$  & $\Delta \log(Z)$ \\
      \hline
      A & 0.36 & 2 & 1.5 \\
      B & 0.18 & 2 & 1.5 \\
      C & 0.36 & 2 & 0 \\
      D & 0.36 & 1 & 1.5 \\
      E & 0.36 & 3 & 1.5 \\
      \hline
    \end{tabular}
  \end{center}
  \caption{\footnotesize List of the simulations: label of the run, power-law index
    of the metallicity dependence by redshift, PopIII yield curves (1,
    2 and 3 refer to lower, central and upper curves of Figure 10 in
    Santoro \& Shull (2006) respectively), allowed variation
    (logarithmically uniformly distributed) in $Z$.  }
    \label{tab:sim}
\end{table}

We now discuss the properties that nuclear clusters possess at birth
and the resulting BH seed population. We start describing our
reference model, A, in Section 3.1 while in Section 3.2 we discuss how
our results depend on PopIII stars metallicity patterns and on the
rate of metal enrichment of the Universe.

\subsection{Model A}

\begin{figure} 
\vspace{0.5cm} 
\includegraphics[width=\columnwidth]{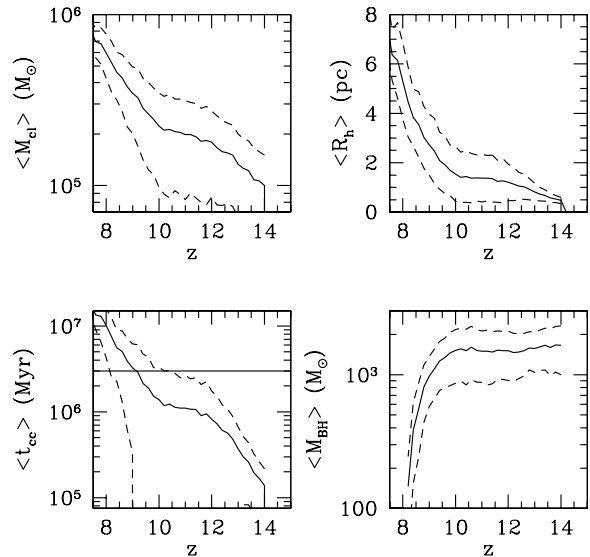}
\caption{Mean cluster masses (upper left panel), radii (upper right),
  core collapse timescales (lower left) and BH masses (lower right) as
  a function of redshift for model A. The orizontal line in the lower
  left panel marks the critical core collapse timescale for VMS
  formation. The dashed curves mark the 1--$\sigma$ scatter. }
\label{fig:fig4} 
\vspace{0.5cm} 
\end{figure}

The first halos reach the critical metallicity for fragmentation at
redshift $\sim 14$. This is also when the first stellar clusters form.
The first clusters have masses of the order of $10^5 \,\msun$ and
$R_h\sim 0.5-1$ pc; in such compact clusters, core collapse starts
early ($\langle t_{cc}\rangle \sim0.1$ Myr at z=14).  At later cosmic
times, the average gas metallicity increases, so that the critical
density for fragmentation decreases.  A lower critical density implies
that $R_{SF}$ increases (Equation ~\ref{eq:RSF}), lengthening the core
collapse timescale.  Therefore, clusters form less concentrated and
more massive and their core collapse timescale continues to increase
with decreasing redshift. This behavior is evident in
Fig. \ref{fig:fig4} for the mean quantities ($ \langle M_{cl} \rangle
$, $ \langle R_{h} \rangle $, $ \langle t_{cc} \rangle $ and $ \langle
M_{\rm BH} \rangle $).  The mean seed mass as a function of redshift
is shown in Figure \ref{fig:fig4} (lower right panel). Unlike $
\langle M_{cl} \rangle $ and $ \langle R_{h} \rangle $, $ \langle
M_{\rm BH} \rangle $ shows no redshift dependence as the increase in
$M_{cl}$ and $t_{cc}$ compensate (see Equation \ref{eq:BH}) leading to
a roughly constant $M_{BH}$.

\begin{figure} 
\vspace{0.5cm} 
\includegraphics[width=\columnwidth]{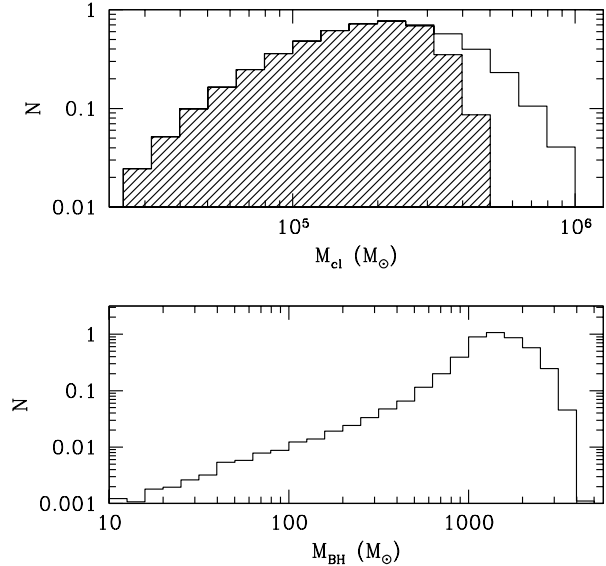}
\caption{Top: Cluster mass function for model A integrated over all
  redshift.  The shaded area corresponds to the subpopulation of
  clusters able to form BH seeds.  Bottom: BH seed mass function
  integrated over all redshifts.  }
\label{fig:fig5} 
\vspace{0.5cm} 
\end{figure}

In Figure \ref{fig:fig5} the entire cluster mass function (solid
histogram) is compared to the mass function of systems able to form a
BH seed (dashed histogram). The clusters that do not form BH seeds are
the very last to form, when the metallicity of the Universe is already
significant.  The BH mass function is also shown in Figure
\ref{fig:fig5}. The mass function is peaked at $\simeq 1000 \msun$
with a long tail at low masses, and a steep drop at high masses.

Fig. \ref{fig:fig8} shows the fraction, $f_{\rm BH}$, of halos hosting
a BH seed. Seeds start to form at $z\sim 14$ in coincidence with the
first, very compact, stellar clusters. The typical core collapse
timescale increases with cosmic time, $\langle t_{cc} \rangle \sim 3$
Myr at z $\sim$ 9, and by $z\sim 8$ $f_{\rm BH}$ drops rapidly.  From
this point on even the most concentrated and least enriched disks are
unable to create central stellar concentrations with $t_{cc}\,<\,3$
Myr and seed formation is completly suppressed. Figure \ref{fig:fig8}
also shows the integrated comoving mass density of seeds $\rho_{\rm
  seed}$. At $z\,\sim\,10$ the mass density saturates at a value of
$\sim\,300\,\msun\,{\rm Mpc}^{-3}$. This mass density should be
considered a lower limit for the total black holes mass density
$\rho_{\rm BH}$ as we are completely neglecting BH growth {\it after}
seeds formation.  We will discuss the implications of this formation
mechanism on the evolution of the supermassive black hole population
in a forthcoming paper. We note however that this seed density is
similar to that expected from PopIII star remnants (roughly a factor
of 3 larger, Volonteri, Haardt \& Madau 2003), and we therefore expect
that most observational constraints can be fulfilled at the same
level.

\subsection{Impact of the uncertainties on metal enrichment onto the seed population}
We now discuss how our results depend on our choice of parameters. We
compute the seed population for BHs formed over cosmic time varying
(i) the history of metal enrichment (models B and C) and (ii) the
ratio of the coolants that determine the $Z_{crit}-n_{crit,Z}$
relationship (models D and E).

Mass functions of protogalactic nuclear clusters for all models
defined in Table 1 are shown in Figure \ref{fig:fig2}. Cluster stellar
masses, $M_{cl}$, span 2 orders of magnitude, between $10^4-10^6
\msun$, with a peak around $\sim\,10^5\,\msun$. Figure \ref{fig:fig2}
also shows the distribution of core collapse timescale of all
clusters. The dashed vertical line at 3 Myr marks the limit for VMS
formation. As discussed in the previous section, clusters satisfying
the condition for the onset of runaway instability are clustered at
small masses and radii: typical masses of runaway unstable clusters
are around a few $10^5\,\msun$ and typical radii are $\sim\,1\,$pc
(see Equation~\ref{eq:tcc}). This naturally points towards the very
first clusters: the first systems that form are indeed those that more
easily can give birth to BH seeds. Consequently, these are also the
most metal-poor clusters, so that our picture is consistent with
requiring that VMS can more easily grow in low metallicity
environments.

As times goes on, both $\langle M_{cl}\rangle$ and $\langle R_h
\rangle$ grow: as a consequence $\langle t_{cc}\rangle $ increases.
Even if clusters continue to form, BH seeds cannot be created any
longer (cfr. Figure~\ref{fig:fig7}). The mass functions of BH seeds
are shown in Figure \ref{fig:fig7}.  The seed mass distributions show
a characteristic shape with a peak at a few $10^3 M_{\odot}$ and a
long tail at lower masses with very little redshift dependence.  This
general picture is valid for all models; we now discuss in turn the
dependencies on specific model parameters.

\begin{figure} 
\vspace{0.5cm} 
\includegraphics[width=\columnwidth]{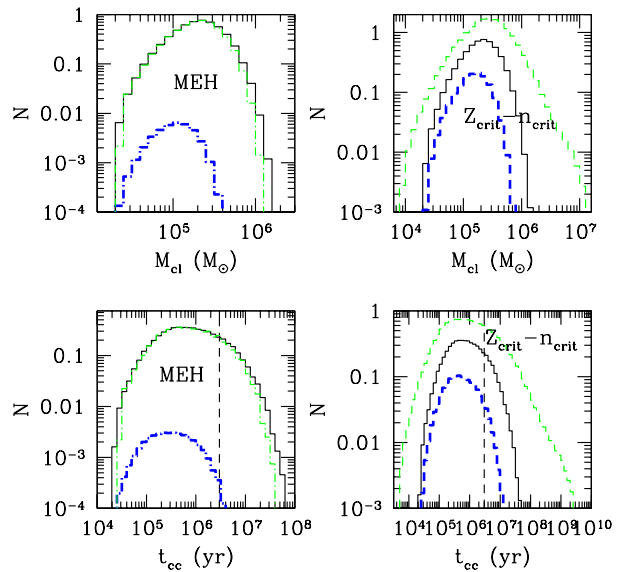}
\caption{Top: Cluster mass functions integrated over all redshift.
  Bottom: distribution of core collapse timescales, integrated over
  all redshifts.  Lines color and style as follow.  Solid: black = A.
  Dotted: thin =B, thick = C. Dashed: thick = D, thin = E.  Model A is
  shown in all panels as a reference.}
\label{fig:fig6} 
\vspace{0.5cm} 
\end{figure}

\subsubsection{Changing MEH}
The MEH is one of the most uncertain parameters. We explore two
extreme cases in models B ($\beta=0.18$, blue dotted curves in Figure
6 and onwards) and C ($\beta=0.36$ and $\Delta \log(Z)=0$, green dotted
curves in Figure 6 and onwards).  The MEH determines, together with
the chosen $Z_{crit}-n_{crit,Z}$ relation, when seeds form.  The
duration of the BH seeds formation epoch is indeed given by a
combination of the assumed metallicity spread, $\Delta \log(Z)$, and of the
slope of the $Z(z)$.  Once $Z$ increases over $Z_{crit,min}$,
fragmentation is activated in a more extended region of the disk, the
inflow is reduced and runaway instability can not proceed
efficiently. The seed formation epoch is therefore longer either if
$Z$ has larger $\Delta \log(Z)$ or if metal enrichment is rather
inefficient.

Model B has the most efficient metal enrichment, and BHs appear
already at redshift 30. On the other hand, BH formation is also
suppressed very early, at $z\,\sim\,18$. At this early cosmic epoch
very few halos were massive enough for efficient atomic line cooling,
thus leading to a comoving seed mass density of only a few $\msun$.
As previously noted for model A $\rho_{seed}$ does not necessary
coincide with $\rho_{\rm BH}$ as we are neglecting seeds growth. In
model B seeds form at higher redshift, and mass can be built up for a
longer period, likely increasing BH masses by accretion.

Model C has the same redshift dependence of the MEH as our reference
model A, but we assume no scatter. A null $\Delta \log(Z)$ (model I)
produces a short burst of seeds very concentrated in time, as seed
formation is allowed only for a very sharp range of $Z$.  Clusters and
BHs form in a burst at $z=11-12$ when $Z\simeq 10^{-5}$ (the minimum
in the $Z_{crit}-n_{crit,Z}$ relation). This burst is very efficient,
with a high fraction of halos undergoing cluster and BH formation, and
the resulting $\rho_{seed}\sim 300 \msun\,Mpc^{-3}$ is similar to our
reference case.

\begin{figure} 
\vspace{0.5cm} 
\includegraphics[width=\columnwidth]{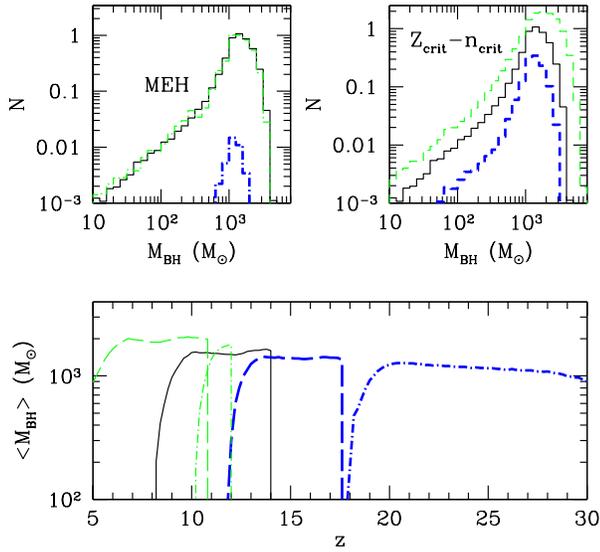}
\caption{Top: Mass function of seed BHs, integrated over all
  redshifts.  Bottom: mean BH mass as a function of redshift.  Lines
  color and style as in Figure~\ref{fig:fig6}. }
\label{fig:fig7} 
\vspace{0.5cm} 
\end{figure}

\subsubsection {Changing $Z_{crit}-n_{crit,Z}$}
We explore the effect of different yield patterns for PopIII stars on
our results in models D (blue long-dashed curves in Figure 6 and
onwards) and E (green long-dashed curves).  The adopted
$Z_{crit}-n_{crit,Z}$ relation defines together with the MEH, when
seeds are born.  The epoch at which BH seeds start to form is related
to the redshift at which the first systems reach the minimum possible
critical metallicity $Z_{crit,min}\simeq 10^{-5}Z_\odot$ (see the
minimum of the curves in Fig. 10 of Santoro \& Shull 2007).

If fragmentation is allowed at lower metallicities (model D) clusters
begin to form earlier.  With increasing cosmic time, $R_{SF}$
increases to $R_{tr}$, thus precluding the formation of a seed. In
model D this happens already at $z\,\sim\,12$, while BH formation
proceeds all the way to $z=5$ in model E.  Consequently, the seed mass
density $\rho_{seed}$ ranges from $\sim\,100$ to 300 $\msun\,Mpc^{-3}$
at $z\,=\,5$, with model D, that forms seed early on, having the lower
seed density.

\begin{figure} 
\vspace{0.5cm} 
\includegraphics[width=\columnwidth]{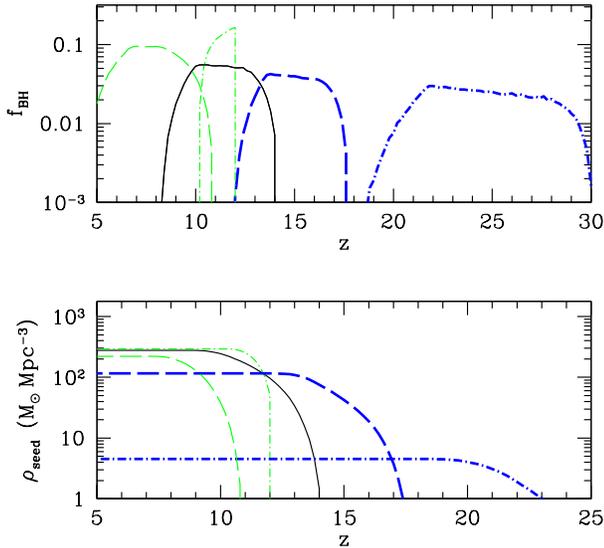}
\caption{Top: fraction of halos hosting a BH seed as a function of
  redshift for models in Table 1.  Bottom: Co-moving mass density of
  the seed BH for the same models.  Colors and lines styles as in
  Figure \ref{fig:fig6}.}
\label{fig:fig8} 
\vspace{0.5cm} 
\end{figure}

\section{Discussion}
We described a model for BH seed formation as result of multiple
successive instabilities.  On a large scale, gravitational torques in
Toomre unstable primordial disks can pile up a significant amount of
gas in the central region of high redshift halos (e.g., Lodato \&
Natarajan 2005).  We focus on relatively massive halos at high
redshift ($T_{\rm vir}\,>\,10^4$ K, $z\gsim 10$) after the very first
stars in the Universe have completed their evolution.  This set of
assumptions ensures that (i) atomic hydrogen cooling can contribute to
the gas cooling process, (ii)  a UV field from the first stars
  has heated the gas, precluding $H_2$ cooling except in the highest
  density regions
, and (iii) the gas inside the halo has been mildly polluted by the
first metals.  The second condition implies that, at least at the low
density at which the collapse starts, $H_2$ is dissociated and does
not contribute to the cooling of the gas (see also Figures 9 and 10 in
O'Shea \& Norman 2008).  The third condition ensures that metals (and
dust) in small quantities can cool down the gas efficiently only as
the gas density reaches a critical threshold. Jointly, the two
conditions ensure that cooling is driven at first by atomic hydrogen
only.

At low metallicity ($Z\sim 10^{-5} Z_\odot$), fragmentation and
low-mass star formation can occur only where the density is above a
metallicity dependent critical threshold ($Z_{crit}-n_{crit,Z}$,
Santoro \& Shull 2006), corresponding to the central densest region of
the protogalaxies.  Within this limited region where star formation
takes place efficiently, very compact stellar clusters form. The
typical stellar masses are of order $10^5\msun$ and the typical half
mass radii $\sim 1$ pc.  Eventually, a large fraction of these very
dense clusters undergo core collapse before stars are able to complete
stellar evolution and a VMS can form (e.g., G{\"u}rkan et al. 2004,
Portegies Zwart et al. 2006).

Clusters unstable to runaway collisions are always the first, less
massive ($\sim 10^5\,\msun$) ones.  As the metallicity of the Universe
increases, the critical density for fragmentation decreases and stars
start to form in the entire protogalactic disk so that i) accretion of
gas in the centre is no longer efficient and ii) the core collapse
timescale increases.  As a result less and less compact clusters form,
and less of them are subject to rapid core collapse.

We computed the properties of the BH population for a set of models,
in dependence of various parameters.  The epoch of seed formation is
determined by the time at which gas in the centre of the halo can
start to fragment. This redshift depends on the metal enrichment
history and on the exact shape of the $Z_{crit}-n_{crit,Z}$
relation. If metal pollution is very efficient and the Universe was
enriched early, the BH formation epoch ends very early, when only a
few halos were massive enough for efficient atomic line cooling.  The
mass and number densities of BH seeds are consequently very low.  The
fraction of halos hosting a BH seed depends also on the fraction of
unstable disk (hence, the critical Toomre parameter, $Q_c$) and on the
frequency of halos with $Z\sim Z_{crit,min}$.  Decreasing $Q_c$ has a
twofold effect on the efficiency of BH seeding.  First, as already
noted by Lodato \& Natarajan (2007), it decreases the number of
bar-unstable disks as the Toomre criterion is satisfied for higher
surface densities (requiring very small spin parameters).
Additionally, the higher surface density for bar-instabilities implies
smaller unstable regions ($R_{tr}$), but a larger $R_{SF}$ at a fixed
particle density. Cluster formation is therefore truncated at
$\lambda<0.004$ for $Q_c=1$, leading to inefficient cluster and BH
formation. The slope of the inner density profile, $\gamma$, affects
$f_{\rm BH}$ as it affects the fraction of runaway unstable systems
that can form a compact cluster.  A lower $\gamma$ produces shallower,
more extended clusters with longer core collapse timescales
(Equations~\ref{eq:RSF} and~\ref{eq:tcc}).  For reasonable choices of
$\gamma>4/3$ we find that the efficiency of the BH formation is within
a factor of 2 of our reference model.

Most of our assumptions have been quite conservative, but still the
population of seeds is comparable to the case of Population III star
remnants discussed, for instance in Volonteri, Haardt \& Madau
(2003). The fraction of high-redshift galaxies seeded with a BH is
about a factor of 10 below the direct collapse case presented in
Volonteri, Lodato \& Natarajan (2008), where a seed was assumed to
form with a 100\% efficiency whenever a protogalaxy disk was Toomre
unstable. An estimate of the degree of agreement between the evolution
of the population of seeds that we have calculated in this paper and
observational constraints requires dedicated models, including the
mass growth of the BHs after their formation, and how galaxy and BH
mergers influence the population. We will present such models in a
future paper.

\section*{Acknowledgements}
We are extremely grateful to Cole Miller for providing insightful
comments on the manuscript.  MV wishes to thank Marc Freitag for
encouraging discussions during the very early stages of this work.

\appendix

\section{Disk and cluster structure}

\subsection{Disk parameters}

Disk parameters $\Sigma_0$, $R_0$, $R_{tr}$ and $m_a$ have been
summarized in Section 2.1. We only insert here the complete expression
for $R_{tr}$ with the $\gamma$ dependency

\begin{equation}
R_{tr}=\frac{2-\gamma}{\gamma-1}\frac{m_aM_h}{2\pi\Sigma_0R_0}.
\end{equation}

As stated in Section 2.1, the particle density of the inner disk
can be written as

\begin{equation}
n(R,h)=n_0\frac{R_0}{R_{tr}}\left(\frac{R_{tr}}{R}\right)^{\gamma}\cosh^{-2}(h/H(R))
\label{eq:nRz}
\end{equation}
\noindent
where $H$ can be computed imposing hydrostatic equilibrium and it is
found out to be $H=\frac{c_s}{\sqrt{2\pi G\mu m_u n_0 f(R)}}$ (see for
example Oh \& Haiman 2002). To simplify the notation we have defined
$f(R)=R_0/R_{tr}(R_{tr}/R)^{\gamma}$. We relate $n_0$ and $\Sigma_0$
by imposing that the surface density calculated starting from Equation
\ref{eq:nRz}, follows the profile described by Equation~3:
\begin{eqnarray}
\Sigma_0 f(R)&=&\mu m_un_0f(R)\int^{+\infty}_{-
  \infty}\cosh^{-2}(z/h){\rm d}z\nonumber \\&=&2 \mu m_un_0f(R) H =
\sqrt{2}c_s\sqrt{\frac{n_0\mu m_u f(R)}{\pi G}}.
\end{eqnarray}

Resolving for $n_0$ we find:
\begin{equation}
n_0=\frac{\Sigma^2_0}{c^2_s}\frac{\pi G f(R)}{2\mu m_u}.
\end{equation}

\subsection{Star formation radius}

Once the density profile of the disk is defined, the star formation
radius can be determined by imposing $n(R=R_{SF},h=0)=n_{crit,Z}$. We
allow clusters to form only if star formation is induced inside $R_{tr}$ and not in the external part of the disk. The
condition for this to happen is that $n_{tr}\equiv
n(R=R_{tr},h=0)<n_{crit,Z}$. $R_{SF}$ is then computed by imposing:

\begin{equation}
n_0\frac{R_0}{R_{tr}}\left(\frac{R_{tr}}{R_{SF}}\right)^{\gamma}=n_{crit,Z}.
\end{equation}

Inserting the expression for $n_0$

\begin{equation}
\frac{\pi G\Sigma^2_0}{2\mu m_u
  c^2_s}\left(\frac{R_0}{R_{tr}}\right)^2\left(\frac{R_{tr}}{R_{SF}}\right)^{2\gamma}=n_{crit,Z}
\end{equation}
 
Resolving for $R_{SF}$

\begin{eqnarray}
R_{SF} & = & R_{tr}\left[\frac{\Sigma_0R_0}{c_sR_{tr}}\sqrt{\frac{\pi
      G}{2\mu m_u n_{crit,Z}}}\right]^{1/\gamma}\nonumber\\ &=&
R^{2/5}_{tr}\left[\frac{\Sigma_0R_0}{c_s}\sqrt{\frac{\pi G}{2\mu m_u
      n_{crit,Z}}}\right]^{3/5}
\label{eq:rsf1}
\end{eqnarray}
\noindent
where in last expression we have inserted $\gamma=5/3$ explicitly.

As stated in Section  2.3, $n_{crit,Z}$ depends on the metallicity
of the gas. The curves in Figure 10 of Santoro \& Shull can be fitted
by the expression

\begin{equation}
\log(Z/Z_{\odot})=a\log^2(n_{crit,Z})+b\log(n_{crit,Z})+c
\label{eq:logZlogn}
\end{equation}
\noindent
where the values of $a,\,b,\,$ and $c$ have been calculated for the
three curves and are reported in Table~\ref{tab:abc}

\begin{table}
  \begin{center}
    \begin{tabular}{|c|c|c|c|c|c|}
      \hline
       $Z_{crit}-n_{crit}$  & 1 & 2 &3 \\
      \hline
      \hline
         a & 0.03517305 & 0.0317305 & 0.0317305 \\
      \hline
         b & -0.582132 & -0.572132 & -0.62132 \\
      \hline
         c & -4.1 & -2.75 & -1.1 \\
      \hline
    \end{tabular}
  \end{center}
  \caption{List of the values of $a$, $b$ and $c$ for fitting different
    $Z_{crit}-n_{crit}$ curves. }
    \label{tab:abc}
\end{table}
\vspace{0.5cm}

The minimal critical density for fragmentation can be computed
from Equation~\ref{eq:logZlogn}. Inserting the expression for the critical density
into Equation \ref{eq:rsf1} one finds

\begin{eqnarray}
\log(R_{SF}) &=& \frac{\gamma
  -1}{\gamma}\log\left(\frac{2-\gamma}{\gamma-1}\frac{m_aM_h}{2\pi\Sigma_0R_0}\right)
+\nonumber
\\&&\frac{1}{\gamma}\left[\log\left(\frac{\Sigma_oR_o}{c_s}\right)+\frac{1}{2}\log\left(\frac{\pi
    G}{2\mu m_u}\right)-\frac{1}{2}\log(n_{crit,Z})\right]\nonumber
\\&=&\frac{\gamma-1}{\gamma}[\log\left(\frac{2-\gamma}{\gamma-1}\right)+A]+\nonumber
\\&&\frac{1}{\gamma}\left(B-C'\sqrt{1+D'\log(Z/Z_{\odot})}\right)
\nonumber
\\&=&\frac{\gamma-1}{\gamma}[\log\left(\frac{2-\gamma}{\gamma-1}\right)+A]+\nonumber
\\&&\frac{1}{\gamma}\left(B-C\sqrt{1+D\beta z}\right)
\end{eqnarray}
\noindent
where in the last expression we have first expressed $n_{crit,Z}$ as a
function of $Z$ and then we have inserted the redshift dependence of
the metallicity\footnote{The last expression refers to the mean $Z$
  only, i.e. it is rigorously appropriate only to model I}$Z$. All
constants except $\gamma$ have been included in $A$, $B$, $C$, $C'$,
$D$ and $D'$:

\begin{eqnarray}
A&=&\log\left(\frac{m_aM_h}{2\pi\Sigma_0R_0}\right) \nonumber\\ B&=&
\frac{b}{4a}+\log\left(\frac{\Sigma_0R_0}{c_s}\right)+\frac{1}{2}\log\left(\frac{\pi
  G}{2\mu
  m_u}\right)\nonumber\\ C'&=&\frac{b}{4a}\sqrt{1-\frac{4ac}{b^2}}
\nonumber\\ C&=& C'\sqrt{1+D'\delta}\nonumber\\ D'&=&
\frac{4a}{b^2-4ac} \nonumber\\D&=& -\frac{D'}{1+D' \delta} \nonumber\\
\end{eqnarray}

Finally for $\gamma=5/3$

\begin{equation}
\log(R_{SF})=k_1+k_2\sqrt{1+k_3\beta z}
\label{eq:rsf2}
\end{equation}
\noindent
where all constants are collected in $k_1\equiv
\frac{2}{5}(A-0.3)+3/5B$, $k_2\equiv 3/5C$ and $k_3\equiv D$.

\subsection{Black hole masses}
Black hole masses depend on the stellar mass of the clusters and on the half-mass radius, via the core collapse timescale.

Cluster masses and timescales to core collapse can be estimated coupling Equations \ref{eq:rsf1} and \ref{eq:rsf2} to Equation 12: 

\begin{eqnarray}
M_{cl}&=&2\pi\epsilon\int^{R_{SF}}_0\Sigma_0R_0R^{\gamma -1}_{tr}
R^{1-\gamma}{\rm d}R\nonumber
\\&=&2\pi\epsilon_{SF}\Sigma_0R_0R^{\gamma
  -1}_{tr}\frac{R^{2-\gamma}_{SF}}{2-\gamma}\nonumber
\\&=&2\pi\epsilon_{SF}\Sigma_0R_0\left[\frac{2-\gamma}{\gamma-1}\frac{m_aM_h}{2\pi\Sigma_0R_0}\right]^{\gamma-1}\frac{R^{2-\gamma}_{SF}}{2-\gamma}
\\&=&\epsilon_{SF}\left[\frac{2\pi}{2-\gamma}\frac{\Sigma^2_0R^2_0}{c_s}\sqrt{\frac{\pi
      G}{2\mu m_u
      n_{crit}}}\right]^{\frac{2-\gamma}{\gamma}}\left(\frac{m_aM_h}{\gamma-1}\right)^{\frac{2\gamma-2}{\gamma}}\nonumber
\\&=&\epsilon_{SF}\left[6\pi\frac{\Sigma^2_0R^2_0}{c_s}\sqrt{\frac{\pi
      G}{2\mu m_u
      n_{crit}}}\right]^{1/5}\left(\frac{3}{2}m_aM_h\right)^{4/5}
\label{eq:massecl}
\end{eqnarray}
\noindent
where in the last line we assume
$\gamma=5/3$.

The core collapse timescale is defined in Equation \ref{eq:tcc}.  The half mass radius can be simply expressed as $R_h=2^{1/(\gamma-1)}\,R_{SF}$.  We use
Equation A11 to express the dependence of $t_{cc}$ on $M_{cl}$ as:

\begin{eqnarray}
t_{cc}&=&\tau_0R^{3/2}_hM^{1/2}_{cl}=\tau_02^{\frac{3}{2}\frac{1}{\gamma-1}}R^{3/2}_{SF}M^{1/2}_{cl}\nonumber=\tau_02^{\frac{3}{2}\frac{1}{\gamma-1}}
\\&\times&\left[\frac{2-\gamma}{2\pi\Sigma_0R_0\epsilon^{1/(2-\gamma)}_{SF}}\right]^{3/2}\left(\frac{m_aM_h}{\gamma-1}\right)^{\frac{3}{2}\frac{1-\gamma}{2-\gamma}}M^{\frac{5-\gamma}{2(2-\gamma)}}_{cl}\nonumber
\\&=& \tau_02^{9/4}\left[\frac{1}{6\pi \Sigma_0
    R_0\epsilon^3_{SF}}\right]^{3/2}\left(\frac{3}{2}m_aM_h\right)^{-3}M^5_{cl}
\label{eq:tccmcl}
\end{eqnarray}
\noindent
where $\tau_0$ is the normalization of Equation \ref{eq:tcc} and its
value in Myr, $\msun$ and pc is 0.019. In the last Equation we assume
$\gamma=5/3$. 

$M_{\rm BH}$ can now be found by inserting Equation \ref{eq:tccmcl} into Equation
\ref{eq:BH}

\begin{eqnarray}
M_{\rm BH} &=&m_*+4\cdot 10^{-3}f_c \ln{\lambda}
M_{cl}\ln{\left(\frac{g}{M^{\frac{5-\gamma}{2(2-\gamma)}}_{cl}}\right)}
\nonumber
\\&\propto& M_{cl}\ln{\frac{\tau}{M^5_{cl}}}
\label{eq:mbhmcl}
\end{eqnarray}
\noindent
where $g$ is defined as

\begin{equation}
g=\frac{t_{\rm
MS}}{\tau_0}2^{-\frac{3}{2}\frac{1}{\gamma-1}}\left[\frac{2-\gamma}{2\pi\epsilon^{2-\gamma}_{SF}\Sigma_0R_0}\right]^{-3/2}\left(\frac{m_aM_h}{\gamma-1}\right)^{-\frac{3}{2}\frac{1-\gamma}{2-\gamma}}.
\end{equation}

From Equation \ref{eq:mbhmcl} we can argue that $M_{\rm BH}$ first
grows with increasing $M_{cl}$, it attains a maximum value and then it
decreases, as more massive clusters have
longer core collapse timescales with $t_{cc}\propto M^5_{cl}$. This last
proportionality results from the dependence of $R_{SF}$ (and
consequently $R_h$) on $M_{cl}$.

\end{document}